# DIFFERENT MECHANISMS OF MACHINE LEARNING AND OPTIMIZATION ALGORITHMS UTILIZED IN INTRUSION DETECTION SYSTEMS

Aziz, Mohammad R. [1, a)] and Alfoudi, Ali Saeed [1,2, b)]

*Author Affiliations*
[1]*University of Al-Qadisiyah, College of Computer Science and Information Technology, Iraq*
[2]*Liverpool John Moores University, College of Computer Science, United Kingdom*

*Author Emails*
[a)] *Corresponding author: mohammad.aziz@qu.edu.iq*
[b)] *a.s.alfoudi@qu.edu.iq*

**Abstract.** Malicious software is an integral part of cybercrime defense. Due to the growing number of malicious attacks and their target sources, detecting and preventing the attack becomes more challenging due to the assault's changing behavior. The bulk of classic malware detection systems is based on statistics, analytic techniques, or machine learning. Virus signature methods are widely used to identify malware. The bulk of anti-malware systems categorizes malware using regular expressions and patterns. While antivirus software is less likely to update its databases to identify and block malware, file features must be updated to detect and prevent newly generated malware. Creating attack signatures requires practically all of a human being's work. The purpose of this study is to undertake a review of the current research on intrusion detection models and the datasets that support them. In this article, we discuss the state-of-the-art, focusing on the strategy that was devised and executed, the dataset that was utilized, the findings, and the assessment that was undertaken. Additionally, the surveyed articles undergo critical analysis and statements in order to give a thorough comparative review. Machine learning and deep learning methods, as well as new classification and feature selection methodologies, are studied and researched. Thus far, each technique has proved the capability of constructing very accurate intrusion detection models. The survey findings reveal that Clearly, the MultiTree and adaptive voting algorithms surpassed all other models in terms of persistency and performance, averaging 99.98 percent accuracy on average.

## INTRODUCTION:

An intrusion detection system (IDS) is software that is capable of monitoring a network or host for possible invasions. Network-Based IDS, Host-Based IDS, and Hybrid IDS are the three most common IDS kinds[1-4]. A network intrusion detection system is located near the firewall (NIDS). It continuously collects and analyzes network data and then takes preventative action[5]. Anti-malware action might be as simple as sending a warning to the administrator or as aggressive as blocking undesirable traffic. When a HIDS detects odd activity in the host's operating system, services, or applications, it generates an alert[6]. Hybrid IDS is capable of operating in both NIDS and HIDS modes[7]. A hybrid intrusion detection system (IDS) receives and analyzes cyber threat data from both NIDS and HIDS. The term "intrusion prevention system" (IDS) refers to an IDS that is capable of preventing assaults (IPS)[8].

According to recent research, there are four basic IDS techniques: It compares incoming signatures to those in its database of known threats by comparing them to statistical patterns indicating normal system activity (e.g., machine learning, biologically inspired systems); it also compares incoming signatures to those in its database of known threats

(e.g., machine learning, knowledge-based systems and multi-agent systems)[9]. Others were less specific. Recent research has also examined IDS tactics in large network infrastructures, notably event stream processing (ESP) systems[10-15].

IDSs fall into three broad categories: network-based IDSs (NIDSs), host-based IDSs (HIDSs), and hybrid IDSs. A NIDS is positioned immediately after the firewall in a demilitarized zone (DMZ)[16]. It records and analyzes network traffic in real-time to take preemptive action[17]. These safeguards against malicious traffic may take the form of individual administrator notifications or proactive measures such as traffic blocking[18]. When HIDSs notice anomalous activity, they examine the host's operating system, services, and installed applications log files and generate alerts[19]. Hybrid IDSs are IDSs that combine the NIDS and HIDS capabilities[19]. A hybrid IDS collects and evaluates heterogeneous cyber threat information using both NIDS and HID[16].

The host-Based IDS approach identifies intrusions caused by the activity of a single host on which it is implemented. It performs a range of checks on various data sources, including recordings of the host's many activities and system log files [1]. This technique identifies malicious activity via the analysis of data generated by the host[20]. The HIDS technique is effective against insider attacks that are not network-related.[2].

Because HIDS monitors just the host, it may detect an intrusion more precisely. Additionally, there is no need for additional hardware or software installation[21]. It is strongly suggested that it be installed on all hosts. Secure communications are simple since they are received by the host and immediately decrypted[8]. Additionally, it must be installed on each host since it consumes resources and increases execution time[21-24].

The IDS, a network-based system, detects and quarantines intrusions and network traffic irregularities. NIDS analyzes the network traffic of a number of linked servers continuously for symptoms of external infection. This approach is good in identifying harmful behaviors perpetrated by outsiders, but it is ineffective at detecting internal threats[25]. NIDS takes advantage of promiscuous network interfaces to examine and filter all traffic passing across a segment of a network and to protect several hosts connected to that segment. When suspicious behavior is spotted in the flow, alarms are triggered[3].

Attacks are identified by the analysis of network data. It is not required to be installed on every host and may operate concurrently on several hosts. The most serious danger in this category is an "Insider attack," which needs highly specialized equipment and is limited to network-based attacks. Encrypted communications are difficult to decrypt, and rapid networks further complicate the task[26, 27].

NIDS provides three types of information in general: signature-based intrusion detection systems (SIDS), anomaly-based intrusion detection systems (AIDS), and hybrid intrusion detection systems (HIDS) (Hybrid IDS).

A signature-based identification system compares the signature on the packet or data to those associated with previously established or ongoing activities. As a result, SIDS refers to previously reported assaults as "Known Attacks." This approach compares the intrusion signature to an existing signature in the signature database, notifies the user of an attack, and raises the alert if a match is discovered[28]. SIDS is more effective against known assaults but fails to detect zero-day attacks because of the database's inability to store new intrusion signatures. The term "knowledge-based detection" is often used to refer to signature-based intrusion detection [4].

AIDS addresses the restriction of not detecting zero-day attacks by monitoring user behavior that deviates from the stated usual user behavior. The term "anomaly" denotes a discrepancy between observed and suggested user behavior. AIDS is not based on the presence of any pre-existing database of signatures.[5]. It identifies intrusions and harmful activities by observing users' abnormal behavior. Due to this method's incapacity to identify undesired activity, a huge number of false alarms may be generated[29]. Numerous research has identified strategies for reducing false alarms[30]. Anomaly detection has a wide variety of applications in various industries, including recognizing medical malpractices, detecting abnormal behavior of moving objects, detecting strange behavior of users on social media platforms, and detecting credit fraud [6].

SIDS is incapable of detecting zero-day attacks, but AIDS may cause a massive number of false alarms; hence, the Hybrid IDS integrates the fundamental ideas of both systems and addresses their deficiencies[6]. A hybrid intrusion detection system comprises several components, including a data collecting module, an analyzer, a signature database, an anomaly detector, a signature generator, and a counter-measure module. Each packet transiting a sensor is read, regardless of whether it is on a single host or a network segment in host-based systems or a network segment in network-based systems. Signatures or regulations are stored in the Signature database. When packets come in quick succession, they are stored in data storage[28, 29]. The Analyzer module then uses a pattern matching approach to compare incoming packets to the signatures or rules stored in the Signature database. The scenario is sent to the anomaly detector if no match is found. The Anomaly Detector performs a scan of the data for discrepancies in pattern recognition. Finally, a message is sent to the signature generator module, which generates and stores a new signature in the Signature Database. This step is repeated for each subsequent signature formed. When the Analyzer detects a match, it notifies the Counter-Measure module and records it in the log file. An administrator may define security policies that indicate which actions should be executed and when. Several authors used this method while building the IDS to increase accuracy, reduce memory requirements, and reduce the number of false positives and negatives [7].

Recently, in light growth of the internet users, this led to the development of an enormous amount of data and the emergence of several types of attacks; this massive amount of data is referred to as Big Data [8].

The growth of computing and social networking data is accelerating, leading to Big Data production. The phrase "Big Data" refers to large amounts of data that are difficult to handle, store, and analyze using conventional database and software approaches. The term "Big Data" refers to a large volume and velocity of data, as well as a diversity of data that necessitates the development of unique management solutions. Additionally, hackers' capabilities have considerably risen. As a result, it is essential to safeguard this massive data gathering. When enormous volumes of data are involved, typical intrusion detection systems become complex and inefficient[26].

As technology advances, cybercriminals enhance their attack strategies, tools, and approaches for exploiting organizations. The public, in particular, has access to public internet services, and many businesses offer their products and services through publicly accessible websites. When an online service is unavailable or compromised, it may negatively impact a business's reputation and income. Generally, security administrators minimize external intrusions by registering all rejected black-list rules for idle services in the firewall. On the other hand, since the Internet Demilitarized Zone (DMZ) is constantly available to the public, firewalls are unable to restrict access to online services. As such, they were discriminating between legitimate and fraudulent access attempts is critical for cybersecurity. Indeed, web-based assaults result in various security events, including data breaches, service disruptions, and malware infection.

The Internet of Things(IoT) is a network of physical components that are supplied with sensors, software, and capabilities to communicate with other devices in the network.. Due to the pervasive nature of these devices and the ease with which they can be monitored and controlled from remote locations, there has been an explosion in the development of novel applications across a variety of domains, including smart homes, wearable devices, health monitoring devices, connected industrial and manufacturing sensors and equipment, and energy management devices. The primary worry with IoT systems is device security and data protection. Cyber assaults are the deliberate exploitation of or illegal access to another individual's or organization's information or infrastructure. Protecting IoT devices from attacks is difficult due to their diversity of devices and protocols, direct internet connectivity, and resource limits. Mitigation strategies designed for IT networks do not apply to the IoT, and only a few Machine Learning models have been built to predict risks based on IoT traffic patterns. Since machine learning algorithms were developed for a variety of applications, including data categorization, clustering, and anomaly detection, they are ideal.

Due to the evolution of cyber threats, data sets used in the studies may become irrelevant to present cyber security issues. Nonetheless, well-known datasets like DARPA and the Knowledge Discovery in Databases (KDD) cup are still being utilized today [9]. The KDD Cup 1999 dataset was utilized in The Third International Knowledge Discovery and Data Mining Tools Competition to develop a prediction model capable of differentiating benign from malignant relationships. On the other hand, it maintains around 5 million connections data, including 41 training features and 24 attacks grouped into four groups [10]: Users-to-Root, Remote-to-Local, and Distributed Denial-of-Service attacks all possible. The KDD cup 1999 was publicly chastised for apparently using fabricated statistics [11]. To begin, simulated data does not represent real-world networks properly. Additionally, it is not a comprehensive account of the assaults (attack behaviors change over time).

Numerous datasets have been suggested to solve the KDD dataset's inadequacies. The first is NSL-KDD, which was created by the University of New Brunswick's Information Security Centre of Excellence[31]. It is similar to the 1999 KDD

Cup, but with a few improvements [10]: Due to the removal of duplicate data, learning models are bias-free. The percentage of records in each difficulty level group is inversely proportional to the percentage of records in the original KDD cup, indicating that classification rates for different learning techniques vary more widely; testing on the entire data set without arbitrarily selecting a small subset is straightforward; and there are not too many records; NSL-KDD does not provide an exact representation of real-world networks [10].

Due to confidentiality concerns, genuine NIDS data sets are private; they include sensitive and private information that cannot be published in the public domain. Additionally, there are other NIDS datasets.[11]. include the Australian Centre for Cyber Security (UNSW-NB15) and KYOTO2016 and the University of New South Wales (ADFA-LD/WD).

The rest of the article is organized as follows. Section II - Different machine learning and optimization algorithms of Intrusion detection. Section III - Different benchmark datasets of intrusion detection … Section VI - Analysis and evaluation research papers. Section V - Conclusion.

# DIFFERENT MACHINE LEARNING AND OPTIMIZATION ALGORITHMS OF INTRUSION DETECTION:

## Machine learning in General Environment

Pu *et al*. [12] addressed the critical problem of anomaly detection techniques requiring regular training on very highly labeled data to identify new threats and reduce false-positive rates (FPA). The authors introduced hybrid clustering, named as (SSC-OCSVM), an anomaly detection approach that combines Sub-Space Clustering (SSC) with One-Class Support Vector Machine (OCSVM). The proposed paradigm is divided into three distinct stages (Initialization, clustering and learning, and Evidence accumulation). The proposed technique tested the model by computing the confusion matrix, recall, FPR, and ROC using the NSL-KDD dataset as a benchmark. According to the authors, each subspace is self-contained and may be processed concurrently, which speeds up execution. The proposed method outperforms both the K-means and the DBSCAN algorithms.

Bhati *et al*. [13] had specified that One of the most challenging aspects of creating an effective IDS is keeping up with the ever-changing and ever-increasing volume of data. There is also a problem with the proportion of FPR that is based on height. They proposed an intrusion detection system based on an ensemble of discriminant classifier schemes with the Random Subspace Algorithm serving as the classifier. Their technique is centered on the development and training of individual classifiers. When classifiers are combined successfully, majority voting-based selections are made. The proposed architecture is organized around four essential stages: data gathering, pre-processing, training, and decision-making. As a result, the framework obtains a high detection rate. On the other side, the authors evaluated their model using ROC, false positives, false negatives, true positives, and true negatives, with KDDcup99 as a benchmark dataset. The model achieved a 98.9 percent accuracy. This model does not account for imbalanced data and explains the results or processing time.

Liu *et al*. [14] have concentrated on the limitations of the traditional SVM when utilized in machine learning-based anomalous intrusion detection. One truth is that some critical categorization information may be lost. The authors presented the ECGC entropy-based clustering strategy, which utilized SVM as local rule models. GA was utilized to optimize the ECGC construction parameters. To a certain extent, the proposed ECGCs might be regarded as extended SVMs. The authors calculated True Positive (TP), False Positive (FP), True Negative (TN), False Negative (FN), and accuracy for their model against the KDDCUP 99 dataset. However, the proposed solution did not account for complexity in its evaluation.

Škrjanc *et al*. [15] Focused on grouping data in a natural environment. These are noisy data that often include outliers, complicating and inaccurate categorization. This work introduced an updated approach for dealing with big noisy datasets and often comprised outliers. It is named evolving Couchiy possibilistic Clustering because it only needs a few tuning parameters, such as maximum density, and is based on cosine similarity. However, this work is entirely theoretical and contains no empirical data or assessment instruments. As a result, no results or evaluations were created.

Meryem *et al*. [16] had addressed the speed of anomaly detection in streaming data with the growth of cloud services. The authors presented a hybrid technique for identifying attacks that incorporates both signatures and 'normal' profile comparison. Additionally, the authors combine anomaly-based and knowledge-based intrusion detection to find unique signatures of attack, dynamically update the training data and critical weight criteria for intrusion detection. The proposed model is divided into five stages (Structuring log events, Eradicating redundancy, Classifying unlabelled Behaviors, Labelling user behaviors and Identifying new signatures of attacks, and updating security rules). These stapeses identified fresh signatures utilizing the Jaccard coefficient, the correlation metric, the Cosine similarity, and security rule updates. Furthermore, the authors assessed the accuracy, precision, recall, and false-positive rate using the NSL-KDD dataset as a benchmark. The best result is reached with KNN, which achieves an accuracy of 98.80 percent, a precision of 99.80 percent, a recall of 98.80 percent, and a false-positive rate of 0.9 percent. While this technique achieves a high detection rate, it falls short when analyzing large-dimensional datasets in terms of consistency.

Another study by Karatas et al. [17] had focused on improving the system's accuracy rate by eliminating the imbalance between classes in the dataset. The authors provide a unique strategy for lowering the imbalance ratio using Synthetic Minority Oversampling (SMOTE). The SMOTE function creates new samples by dividing the feature vector by its nearest neighbor and then multiplying the result by a random value between 0 and 1. The result is then appended to the feature vector that was analyzed. To begin, pre-process the data by filling in missing values, setting the maximum value to infinity value, dividing the Timestamp feature into two columns, digitizing and normalizing the data, and then applying six machine learning algorithms sequentially on the pre-processed data (DT, K-NN, Gradient Boosting, RF, Adaboost, and Linear Discriminant Analysis algorithms). Five attack classes must be selected, plus one for non-attack. Then, evaluate the performance of these algorithms. To summarize, we will analyze the performance of multiple machine learning methods by first decreasing the dataset imbalance ratio via the use of (SMOTE). The Accuracy, Precision, Recall, F1-Score, and Error Rate variables are evaluated when applied to the KDD-CUP99 dataset, the NSL-KDD dataset, the CIC-IDS2017 dataset, and the CSE-CIC-IDS2018 dataset. This experiment achieved outstanding results, with a precision of 99.69 percent, a recall of 99.34 percent, an F1-Score of 99.35 percent, an error rate of 0.65 percent, and a time of 274 seconds. Indeed, the analysis of six machine learning-based NIDSs focuses on resolving imbalanced dataset and Reducing the level of imbalance via the application of SMOTE. This finally improves the detection rate of minority attack classes, as measured by the most current dataset CSECIC-IDS2018. While the Adaboost approach enhances detection accuracy, the cost is higher execution time. Additionally, this study might have delved farther into data purification.

Yin et al. [18] had focused on poor detection accuracy and instability for small categories of data in NIDs. The authors proposed an expanded Bayesian network (KDBN) structural model for networks and utilized the (MAP) criterion to develop an IDCM based on the enhanced KDBN. Then it pioneered the virtual augmentation strategy to improve the efficiency, accuracy, and stability of NIDs. The proposed technique determined the model's accuracy by comparing it to the KDDCup99 dataset. Additionally, the accuracy rate was greater than 95%. Accuracy alone will not be enough for this investigation's evaluation.

Jiang et al. [19] had addressed that improving NIDS detection accuracy is an essential subject in cybersecurity. This article introduces the PSO-Xgboost model based on Xgboost. The PSO algorithm utilizes to optimize its parameters adaptively rapidly. Xgboost supports linear classifiers and includes the second derivative in the Taylor expansion of the cost function. The optimal structure of Xgboost was discovered in this study utilizing PSO. When using the NSL-KDD dataset as a benchmark, the authors assessed the precision, recall, and F-Score values. This strategy was effective in achieving beneficial results. The precision is 0.81, the recall is 0.75, and the F-Score is 0.71. This experiment made no use of an algorithm, processing time, or accuracy rate. Due to the considerable calculation involved, this strategy may not provide the optimal solution worldwide. As a result of the results, we may determine that data is duplicated, not adequately centralized, and wrongly classified, such as U2R and R2L.

This study achieved by Uhm et al. [20] had focused on improving the performance of classification in terms of accuracy and speed and developing robust and reliable NIDSs working in various environments. This article presented a novel method for segmenting datasets according to their service awareness. The primary approach is to break down extensive data into smaller parts. The breakdown's objective is to decrease the total number of service categories and related attacks in each sub-dataset. The classifier proposed in this article is composed of two partitioning components ( static and dynamic divisive partitioning). The first one eliminates redundant data and organizes it into smaller sub-datasets based on service, destination port, and session flag. The second strategy is to mix agglomerative and divisive partitioning in order to reduce the attack type size for each partition. When an RF classifier is utilized, this significantly enhances classification performance. The Kyoto2016 dataset for binary classification and the CIC-IDS2017 dataset for multi-classification were utilized in this proposal. Additionally, evaluation metrics such as Accuracy, Precision, Recall, and Fl-score are computed. A critical point was made in this suggestion about the use of datasets and the performance of classifiers in Similar to the natural environment, different datasets were utilized in a variety of ways and the effects were closely analyzed, but there was no clearly defined evaluation technique used to evaluate the recommended strategy. Furthermore, no extra thorough data preparation was conducted.

## Deep learning in a general environment

Gao *et al*. [21] had focused on detecting malicious attacks in the data with a small proportion. This paper established an adaptive ensemble learning model (using MultiTree and adaptive voting algorithms) that combines the advantages of each technique for different types of data detection and achieves optimal results through ensemble learning. The voting mechanism determined which detection performance had the highest likelihood of being detected. Ensemble learning provides the advantage of integrating the predictions of a large number of base estimators to improve generalizability and robustness compared to a single estimator. SVMs (support vector machines), logical regression, kNNs (k-nearest neighbors), Adaboost, random forest, and deep neural networks are all alternative classifiers. Using the PCA principle component analysis technique, they evaluated five classifiers for the voting system's composition. The authors utilized the NSL-KDD data set as a benchmark dataset. Additionally, the proposed model demonstrated an accuracy of 85.2%. In this model, accuracy alone is insufficient for evaluation; the false positive rate, F1 score, and precision are essential factors.

Lv *et al.* [22] had addressed developing an accurate and effective misuse intrusion detection system that utilizes machine learning methods and relies on particular attack signatures to accurately and rapidly differentiate between regular and malicious activity. To optimize the parameters of the model, the authors introduced (KPCA-DEGSA-HKELM), a model based on an Extreme Learning Machine and a hybrid kernel function developed from Differential Evolution (DE) and Gravitational Search Algorithms (GSA). Kernel principal component analysis (KPCA) reduces the dimension of nonlinear identification data and extracts features. In reality, this methodology comprises the following five steps: (collection of data and extraction of sequences, extraction and processing of features, clustering, extraction of behavioral sequences, detection and classification). Precision, Recall, and F-scores are used to assess each class. The proposed DEGSA model was evaluated against the KDD99 and UNSW-NB15 datasets and achieved an accuracy of 96.69 and 89.01 percent, respectively, when accuracy, mean accuracy, mean F- score, attack accuracy, false alarm rate, and false normal rate were used as assessment metrics. This method is capable of detecting records in Difficult that is not irregular.

Yu *et al.* [23] had focused on learning the network intrusion detection model from a small amount of labeled data. Because the normal class data volume is often much larger than the abnormal sample volume, the proportion of each category in the abnormal samples is very disparate. The authors proposed an intrusion detection model using FSL to identify abnormal network activities and a balanced resampling technique to maintain the dataset's balance throughout training. Two network models (DNN) were used to create the FSL (CNN). The performance of the proposed model is improved by the inclusion of the centerless function often used in face recognition. The proposed model needs training on less than 2% of labeled data. Three key elements impact the classification results achieved using FSL: embedding, distance, and loss function (used to boost accuracy). The evaluation based on a similarity metric (ACC, PRE, REC, F-M, FAR). The benchmark datasets NSL-KDD and UNSW-NB15 were utilized. On UNSW-NB15, the proposed model achieved accuracy rates of 92.34 percent and 85.75 percent for KDDTest+ and KDDTest+22 FAR, respectively (8.01 percent). Indeed, the model achieves decent detection rates for U2R and R2L attacks on the NSL-KDD dataset, while these rates are still lower than those for other attack classes in the dataset. Additionally, FSL needs labeled data for learning, which restricts its use in situations where the model requires frequent training on unlabeled network traffic to aid it in learning additional patterns for effective detection, and FAR was not a suitable output. The key downside of the technique is that FSL requires a balanced dataset to uncover anomalies for intrusion detection.

Su *et al.*[24] were had addressed the intrusion detection can be considered a classification problem, intrusion detection accuracy should be improved. The authors defined BAT as a strategy for improving feature selection and accuracy in IDS classification by merging bidirectional long-term memory (BLSTM) and attention processes. The suggested technique consists of four stages (Data Preprocessing Layer, Multiple Convolutional Layers, BLSTM Layer and Attention Layer). Using the single-hot approach, the data were digitized and standardized. A convolutional layer was used to extract local characteristics from traffic data. By linking forward and backward LSTM, the BLSTM model captures coarse-grained information. Because the attention mechanism was employed to modify the likelihood of packet vectors, this model prioritizes key traits. Back-propagation through time (BPTT) is used to calculate the error differentials, the derivatives of the objective function concerning all weights are found, and the objective function is minimized using stochastic gradient descent. The softmax classifier was used to classify network traffic. The authors use the NSL-KDD dataset to determine Accuracy, False Positive, false-positive rate, True Positive, true-positive rate, False Negative, and false-negative rate. Around 84.25 percent of the time, it was correct. While the suggested approach improves efficiency, ANNs are slow to train and prone to local minima.

# Feature selection

Almomani [25] had addressed that the feature selection(FS) approach reduces the processing time of traffic behavior and increases accuracy. The authors described BAT as a strategy for improving IDS classification accuracy and feature selection by combining BLSTM and attention processes. The suggested technique has four phases (Data Preprocessing Layer, Multiple Convolutional Layers, BLSTM Layer and Attention Layer). The data were digitized and normalized in one step. A convolutional layer collected local information from traffic data. It connects forward and backward LSTM to retrieve coarse-grained information. Because the attention mechanism was utilized to change packet vector probability, this model prioritizes key traits. The objective function's derivatives concerning all weights are found, and the objective function is minimized via stochastic gradient descent. The softmax classifier classified network traffic. Their evaluation measures include Accuracy, False Positive, False Positive rate, True Negative, and False Negative rates utilizing the NSL-KDD dataset. It was 84.25% correct. It is important to note that ANNs are sluggish learners and are prone to local minima.

Shunmugapriya *et al.* [26] had focused speed on Feature Selection (FS) to improve the classification process. FS is seen as an optimization problem because selecting the appropriate feature subset is very important. To overcome ACO stagnation and ABC delayed convergence, this article provides an AC-ABC Hybrid Algorithm. ACO generated potential solutions based on its pheromone values. The generated solutions were then utilized as food sources by three distinct bee species present in ABC. ACO's pheromone was changed with ABC's most incredible meal. Each cycle was repeated until a certain number was reached. The proposed hybrid algorithm beat both ABC and ACO. However, the benchmark datasets have few characteristics (less than 100). The proposed technique also seems to have a low convergence rate, given the large maximum number of iterations (1000). The ant system examines the dataset and returns the optimal feature subset combination. It shows the colony's early food sources. The EBees employ feature subsets to increase prediction accuracy. Then each feature subset's goal function and fitness values are calculated. The OBees chose the feature subset that gives the best-projected outcomes based on fitness. The observer bee's created feature subset determines the best ant and global pheromone update. The ants' newly produced feature subsets are given to the bee colony to use as food sources. The algorithm ends when the number of iterations is achieved. The suggested approach yields a feature subset with the maximum possible accuracy. For the categorization power level, we used the F-Test (MANOVA). The dataset is made up of several. Cardiology, dermatology, and other fields. Die empfohlene Antwort was 99.43 % correct. This strategy's main flaw is that it starts with filtering, which might remove valuable features.

Malik *et al.* [27] had focused on improving the false-positive rate(FPR) in Anomaly detection systems. This paper presented a hybrid strategy that used the binary particle swarm optimization algorithm for pruning the decision trees. Then use PSO to trim the tree by competing with all nodes except the root and leaves. It starts with a random binary bit-string of 1s and 0s (chosen) (dropped). The number of 1s and 0s in each particle equals the number of UDT branch nodes. Swarm particles each have a pruned decision tree. The sigmoid function was employed to check for a certain branch node in the tree. It prunes random branch nodes one at a time to see whether it improves classification accuracy. To categorize network intrusions, PSO's pruned decision tree was utilized to classify them using two previously suggested approaches: single-objective optimized decision tree pruning (SO-DTP) and multi-objective optimal decision tree pruning (MO-DTP) (FPR). The MO-DTP algorithm tries to balance both goals by guaranteeing that both are met to some degree and that one does not outperform the other. VEPSO is a strategy for assessment based on the vector evaluated genetic algorithm (VEGA). The evaluator's The KDD99Cup dataset was used to compute IDR, FPR, classification accuracy, precision, tree size, classification time, and cost. MO-DTP employed IDR and FPR to optimize a decision tree. The proposal obtains an IDR of 92.713 percent with a 0.033 class cost. Accuracy and precision are 96.66% and 99.98% for this proposal with an accuracy of 96.65 percent and 99.98% for the precision. Uncertainty about the projected forecast may make achieving the global optimum solution within the time restrictions. Second, the PSO approach's traditional global and personal best upgrading processes may lose certain essential aspects.

Gu *et al.* [28] had focused on reconstructing the original data on the original data to obtain high-qualified training data for improving detection accuracy and increasing training speed. This article presented a paradigm for novel intrusion detection based on feature-enhanced SVM ensembles by applying fuzzy c-means clustering to group the raw data. Following that, new data is generated using ratio transformations, and finally, SVM classifiers are trained in the bottom layer, and their outputs are delivered to the top layer SVM. The NSL-KDD dataset was used as a benchmark. Accuracy, detection rate (DR), false alarm rate (FAR), confusion matrix, true positive, true negative, false negative (FN), and false-positive were then used as performance measures for the proposed model. NSL-KDD is the dataset. This model achieved an accuracy of 99.41 percent, a detection rate of 97 percent, a false alarm rate of 0.4 percent, and a training time of 2.6 seconds. This study does not contain the pseudo-code for an algorithm.

Nazir *et al.* [29]. Had focused on reducing the dimensionality of network traffic data to improve IDS in a real-time environment. The authors proposed Tabu Search - Random Forest (TS-RF) as a unique wrapper-based feature selection approach that makes use of a weighted-sum-based fitness function. The TS algorithm is used to do property searches, while the RF approach is used to perform learning. The authors compared TS-RF to three well-established feature selection techniques (Gain Ratio, Chi-Square, and Pearson Correlation). Furthermore, they said that their technique maximizes the model's classification accuracy and temporal complexity. Their fitness function considers three criteria: classification accuracy, false-positive rate, and feature count. This information was used to assess the technique's overall effectiveness. Additionally, using the UNSW-NB15 dataset, the authors construct the accuracy, false-positive rate, and Receiver Operating Characteristic (ROC) values for assessment. The proposed model achieved an accuracy of 83.12% and an FPR of 3.70%. Indeed, the authors did not believe the data were unequal, and they entered the initial values for their suggested parameters manually.

Halim *et al.* [30] had focused on addressed here to devise an improved unsupervised feature selection technique to improve detection accuracy with high True Positive TP and low False Positive FP. To improve the accuracy of classifiers, the authors suggested an upgraded Genetic Algorithm (GA)-based feature selection approach called (GbFS). Firstly. Data normalization is the process of transforming data into a more manageable format. Then, improved GA with fitness function was applied using two processes (objective function and scaling function) and the roulette wheel selection strategy, which has the benefit of reducing execution time and requiring no scaling/sorting, as other selection techniques do. Then, categorizing (SVM, k-NN, XgBoost). Accuracy and recall are the assessment criteria.

## Real-time environment

Sangkatsanee *et al.* [31] had focused on reducing The number of features to improve the complexity of computation and the amount of computer resources required. The authors suggested a method for detecting intrusions in real-time (RT-IDS). Three separate steps comprise the proposed paradigm. (Initialization, Classification, and Post-processing). Twelve characteristics were extracted from the header data of network packets using a packet sniffer. The primary features have been linked to the data collection process. Categorization was accomplished via the usage of a Decision Tree method. Post-processing is used to clean up the classification result by removing outliers or false alarm detections. We propose that for every five consecutive detection results for each pair of IP Addresses (source and destination), a majority voting method be used to determine whether the result reflects normal network activity or an attack type. A majority vote approach was used in the post-processing stage to eliminate outliers or false alarm detections from the categorization result. Each pair of IP Addresses (source and destination pair) had its five consecutive detection findings evaluated using a majority vote approach to determine if the result represented regular network activity or an attack type. The authors assessed the KDD99 dataset as well as an RLD09 dataset collected in a real-world network environment. Additionally, we calculated the total detection rate (TDR), the normal detection rate (NDR), the DoS detection rate (DDR), and the probe detection rate as assessment metrics (PDR).

Seo *et al.* [32] had focused on developing a quick and accurate algorithm for real-time attack detection with high attack detection accuracy. The authors argued for the implementation of a two-tier intrusion detection system. Utilizing packet- and flow-based classifiers, it enhances performance and accuracy. The level 1 classifier starts by extracting a subset of the packet's properties to enable rapid classification, allowing real-time attack detection. The level 2 classifier can only identify flows that were classified incorrectly by the level 1 classifier.

As a consequence, a time-consuming machine-learning-based classifier can control the traffic. Due to the unique structure of the two-level classifier, classification speed and accuracy may coexist. The authors assessed the datasets UNSW-NB15 and CICIDS2017, calculating accuracy, precision, recall, and F-score, and finding that the datasets had an accuracy of 98.1 percent and 99.98 percent, respectively. Due to the timing constraints of this study, a solution for constructing real-time intrusion prevention systems (IPSs) capable of overcoming the inadequacies of present network security is required. This proposal creates an implementation structure without a stated method, meaning that no specific result would be achieved. It compares several methods by using the vinyl as the evaluation measure rather than the processing time.

H. Zhang *et al.* [33] aimed at the slow detection speed for massive data. The authors introduced a SC&D-S intrusion detection model based on DBN-SVM. They make full advantage of deep learning's capabilities for lowering the dimensionality of large nonlinear data sets with a high dimension. The technique builds a pattern library from regular patterns, swiftly discovers abnormalities in the data stream, and updates the pattern library incrementally. Then, using a

DBN deep learning technique based on multilayer nonlinear mapping, the feature dimensions of high-dimensional and nonlinear original data are minimized, reducing the number of feature vectors while maintaining essential data features. Then, one-to-all classification is conducted using multiple SVM classifiers; multiple classifications are generated using the voting method; and intrusion identification is performed on the low-dimensional feature vectors collected using the DBN approach. The DBN and SVM training techniques are utilized separately in this model. As evaluation criteria, total time, precision, recall, and F1-score were used. The authors benchmarked their method against the CICIDS201 dataset and found a precision of 0.93920. Accuracy was assessed in this study by an examination of the outcomes.

Kim *et al.* [34] focused on the inefficiency of intrusion detection systems due to false-positive alarm events and not adaptively to the natural world environment in real-time. The authors proposed a CNN-LSTM model for spatial feature learning (SFL). The suggested technique is divided into three components: data collection and splitting in order to acquire online traffic and split training data for each model, data preparation and training utilizing labeled analytical information, and finally, prediction for suspicious payloads on fresh web traffic. Accuracy, recall, confusion matrix, precision, and F-Score were used as evaluation criteria. We used the benchmark datasets CSIC-2010 and CICIDS2017. The proposed strategy achieved an accuracy of 99.99 percent.

## Fog environment

Onah *et al.* [35] had focused on implementing an Intrusion detection system model over massive datasets in the fog computing. This study created a Genetic Algorithm Wrapper-Based Feature Selection and Nave Bayes for Anomaly Detection Model (GANBADM) in a fog environment that eliminates irrelevant features while retaining excellent accuracy. For classification, GA was used as a random search technique with the Naive Bayes classifier. Precision, Accuracy, F1 Score, and Execution Time had all become performance indicators. The dataset was utilized as a benchmark by the NSL-KDD. The proposed approach achieved a 98 percent True-Positive Rate, a 0.6% False-Positive Rate, and a 99.73 percent accuracy.

Khater *et al.* [36] focused on the security issue in fog and IoT devices due to resource limitations. The authors proposed a light fog computing intrusion detection model based on a single hidden layer Multilayer Perceptron (MLP). To extract features, an N-gram transformation was utilized. Additionally, the matrix formatting is compressed using a sparse matrix. Additionally, the linear correlation coefficient (LCC) is used to adjust for data with zero values. Previously, the number of characteristics was limited by selecting features based on mutual information. The recall, F-Measure, accuracy, CPU use, testing duration, and energy consumption were all evaluated. The authors apply their model to the Australian Defense Force Academy Linux Dataset (ADFA-LD) and the Australian Defense Force Academy Windows Dataset (ADFA-WD), achieving 94% accuracy, 95% recall, and 92% F1-Measure in ADFA-LD and 74% accuracy, 74% recall, and 74% F1-Measure in ADFA-WD, respectively. Indeed, due to the result's invitation, the suggested technique requires a substantial amount of detection time in contrast to fog computing[26].

De Souza *et al.* [37] had focused on the lack of robust intrusion detection techniques that can guarantee a safe and suitable environment for applications based on IoT. Khater et al. [43] concentrated on the security problem that arises in fog and IoT devices due to resource constraints[21].

Sadaf *et al.* [38] focused on intrusion detection systems for the Fog environment. The authors proposed the term (Auto-IF) to refer to an approach that combines autoencoder (AE) and isolation forest (IF) methodologies[3, 31, 32]. It detects irregularities in two stages. AE detects the assault and distinguishes it from regular network traffic. They then submit the generated sets to the Isolation forest, which detects and eliminates outlier data points, improving overall accuracy. The training and testing datasets were normalized after pre-processing. The proposed approach would take into account all of the data collection attributes. The technique was evaluated using the KDD CUP99 and NSL-KDD datasets. The authors validated their approach against the NSL-KDD benchmark dataset and discovered that it achieved a 95.4 percent accuracy rate, a 94.81 percent precision rate, a 97.25 percent recall rate, and a 96.01 percent F-measure.

An *et al.* [39] focused on security precautions to effectively handle security threats in FC infrastructure and minimize the associated damage. The authors proposed the SS-ELM algorithm, a distributed ELM classification algorithm for fog networks that combines sample selection and Extreme learning machine (ELM) according to the FC/MEC network characteristics, in which each fog node is trained on a sample of the entire data with the assumption that this sample represents accessible data to the fog node[16, 33]. The authors established a typical ELM model by indexing it against the fog node and requested that the training method be repeated until a minimum necessary performance of training error, referred to as a performance index, was attained. Select training samples depending on the training properties of the network and each fog node/MEC host. Six steps comprise the proposed strategy (initialize the sample sets and network structure parameters, create random hidden layer node parameters, calculate the hidden layer output). The accuracy and false positive rate were determined. The KDD Cup99 dataset was used as a benchmark, which had an accuracy of 99.07 percent and a training time of 4.52 seconds[34, 35].

# Other studies.

Somasundaram *et al.* [40] efforts have been directed at establishing an efficient classification model for automated diagnosis of diabetic retinopathy diseases. First, by using the bagging technique to extract critical characteristics through ML-FE and t-SNE, we can construct a probability distribution across pairs of objects to produce similar items with a higher possibility of being chosen and vice versa. Then, by comparing the (higher and lower) dimension counterparts and using the Kullback Leibler divergence to exclude the lower dimension counterparts, extracting features becomes more straightforward without compromising their edges. Second, simultaneously and separately train the proposed BEC model on the attributes with equal weighting for each. The results are then aggregated using the majority voting technique. The system was evaluated using the following metrics: Sensitivity, True-Positive, False-Negative, Classification Time, and Detection Rate. The Standard Diabetic Retinopathy Database (DIARETDB1) was used as a benchmark dataset.

Zhang *et al.* [41] had focused on detecting anomaly network performance based on machine learning algorithms. The authors developed two innovative techniques for detecting network performance anomalies using split-sample classification: AdaBoost and a simple feedforward neural network. By separating data into subgroups based on a particular characteristic, boosted decision trees train a decision tree. The decision tree is next evaluated against the training data to detect misclassified data values. Each value is assigned a weight that reflects its importance. A new tree is built when the weights of wrongly classified items increase and those of correctly classified values decrease. It then compares the original and freshly created trees to the training data to discover misclassified values. Another tree was constructed, and so on until the required number of weak classifiers, i.e., estimators, was acquired. Each tree will yield a weighted vote of 0 or 1 to indicate whether or not anything is unusual. The majority vote was used to categorize datasets, and the accuracy and AUC scores were used to evaluate. We developed a simulation of real-world data and collected real-world data from several networks to establish their one-way latency, throughput, and packet loss. This data was used to verify the efficacy of our tactics. This study contains no processing time or assessment statistic, such as the False-Positive rate[36].

Vijayanand *et al.* [42] had addressed the feature selection approach in the intrusion detection system. The authors combined a GA-based wrapper strategy with an SVM Classifier to build a GA-based wrapper approach. To begin, normalize the binary output of multiple class training data. To ensure that data is dispersed equally. It is possible to identify the informative qualities of a given class by considering the output of the associated class as (1) and the output of all other classes as (0). Second, by use a GA to extract significant features about each attack type for the SVM classifier. Each attack category is assigned its own classifier, which is trained in the second step using the informative features retrieved from the attack data. A SVM classifier was used to assess the fitness values. The RBF kernel function is used to determine the maximum margin of the SVM classifier and has a sufficiently fast convergence rate. The words Accuracy, FPR, FNR, Sensitivity, Specificity, and precision were used throughout the review process. As benchmarks, the ADFA-LD and CICIDS2017 datasets were employed, which obtained 99.95 percent and 99.39 percent accuracy, respectively[29, 34, 35, 37-40].

Alfoudi *et al.* [43] had focused on building a feature reduction and selection method to increase the quality of the generated features of Palm Vein Identification. The authors suggested a combination model of two-dimensional discrete wavelet transform, principal component analysis (PCA), and particle swarm optimization (PSO) termed (2D-DWTPP), which feeds an optimum subset of data to the wrapper model in order to increase palm vein pattern prediction accuracy. The 2D-DWT collects features from palm vein images, reducing palm vein feature redundancy through PCA. To begin, we will apply adaptive histogram equalization to the grayscaled image to enhance its quality and then transform the resultant image to a negative image as a pre-processing step. The pre-processed image is then represented as a succession of low pass and high pass apexes using two-level DWT (high and low pass), followed by PSA to reduce the number of big and redundant features formed by the subsequent phase. Finally, PSO is utilized to accelerate the process of determining the optimal attributes. The authors used a support vector machine classifier to categorize the model's properties. Accuracy was used as an evaluation criteria. The dataset contains data that the researcher acquired manually. As a result, the accuracy rating is 98.65%[41].

# ANALYSIS AND EVALUATION OF RESEARCH PAPERS

To assess the outcomes of the intrusion detection methodologies developed by various researchers via this survey, and to provide an accurate overview of all the articles included, the primary criteria to be explored and compared is the exhibited accuracy[14]. Indeed, the majority of study articles focused on accuracy measures[37, 38]. In practice, the accuracy measure is a critical component in evaluating machine learning system[20, 42, 43].

We have classified 37 approaches in this literature into eight groups based on their environment. All results are depicted in eight tables.

**Table 1** shows resultant accuracy by Machine learning in General Environment in [12] , [13] , [14] [42], [15] , [16] , [17] , [18] , [19] , [20] and the study of Karatas et al. achieved highest accuracy. **Table 2** shows resultant accuracy by Deep learning in a general environment in [21] , [22] , [23] ,[37, 44] [24] [45]and the study of Lv et al. achieved highest accuracy. **Table 2Table 3Table 2** shows resultant accuracy by Feature selection in [25], [26] , [27] , [28] , [29] , [30] and the study of Halim et al. achieved highest accuracy. **Table 4** shows resultant accuracy by Real-time environment in [31], [32],  [33], [34] and the study of Kim et al. achieved highest accuracy. **Table 5**  shows resultant accuracy by Fog environment in  [35], [36], [37], [38], [39] and the study of De Souza et al. achieved highest accuracy. **Table 6**  shows resultant accuracy by Other studies. in [40], [41], [42], [43] and the study of Vijayanand et al.  achieved highest accuracy.

Table 1. Machine learning in General Environment

| Article | Proposed solution | Dataset | Accuracy (%) |
|---|---|---|---|
| Pu et al. [12] | (SSC-OCSVM) | NSL-KDD | |
| Bhati et al. [13] | Ensemble-based technique | KDDcup99 | 98.9 |
| Liu et al. [14] | ECGC | KDDcup99 | |
| Škrjanc et al. [15] | eCauchy clustering based on cosine similarity | DARPA | - |
| Meryem et al. [16] | Hybrid intrusion detection system using machine learning | NSL-KDD | 98.80 |
| Karatas et al. [17] | SMOTE | KDD CUP99 NSL-KDD CIC-IDS2017 CSE-CIC-IDS2018 | 99.69 |
| Yin et al. [18] | KDBN | KDDcup99 | 95 |
| Jiang et al. [19] | XGBoost | NSL-KDD | |
| Uhm et al. [20] | Two-level partitioning algorithm | KYOTO2016 CIC-IDS2017 | 98.74 99.98 |

Table 2. Deep learning in a general environment

| Article | Proposed solution | Dataset | Accuracy (%) |
|---|---|---|---|
| Gao et al. [21] | ensemble voting approach | NSL-KDD | 85.2 |
| Lv et al. [22] | KPCA-DEGSA-HKELM | KDDcup99 UNSW-NB15 | 96.69 89.01 |
| Yu et al. [23] | Few-Shot Learning | NSL-KDD UNSW-NB15 | 92.34 92.00 |
| Su et al.[24] | BAT | NSL-KDD | 84.25 |

Table 3. Feature selection

| Article | Proposed solution | Dataset | Accuracy (%) |
|---|---|---|---|
| Almomani [25] | PSO,GWO,FFA and GA based FS | UNSW-NB15 | 90.484 |

| Article | Proposed solution | Dataset | Accuracy (%) |
|---|---|---|---|
| Shunmugapriya et al. [26] | AC-ABC | Wisconsin | 99.43 |
| Malik et al. [27] | Prune DT using PSO | KDDcup99 | 96.65 |
| Gu et al. [28] | FS based on SVM ensemble | NSL-KDD | 99.41 |
| Nazir et al. [29] | TS-RF | UNSW-NB15 | 83.12 |
| Halim et al. [30] | GA based feature selection | CIRA-CIC-DOHBrw-2020 UNSW-NB15 Bot-IoT | 99.80 |

**Table 4.** Real-time environment

| Article | Proposed solution | Dataset | Accuracy (%) |
|---|---|---|---|
| Sangkatsanee et al. [31] | RT-IDS | KDDcup99 RLD09 | |
| Seo et al. [32] | Two-level IDS in real time | UNSW-NB15 CIC-IDS2017 | 98.1 99.98 |
| H. Zhang et al. [33] | SC&D-S | CIC-IDS2017 | 93.92 |
| Kim et al. [34] | CNN-LSTM | CSIC-2010 CIC-IDS2017 | 99.9 |

**Table 5.** Fog environment

| Article | Proposed solution | Dataset | Accuracy (%) |
|---|---|---|---|
| Onah et al. [35] | GANBADM | NSL-KDD | 99.73 |
| Khater et al. [36] | MLP based Lightwight ID model | ADFA-LD ADFA-WD | 94 |
| De Souza et al. [37] | DNN-KNN | NSL-KDD CIC-IDS2017 | 99.77 99.85 |
| Sadaf et al. [38] | Auto-IF | KDD CUP99 NSL-KDD | 95.4 |
| An et al. [39] | SS-ELM | KDD CUP99 | 99.07 |

**Table 6.** Other studies.

| Article | Proposed solution | Dataset | Accuracy |
|---|---|---|---|
| Somasundaram et al. [40] | ML and ensemble clasifier using rubustFE | DIARETDB1 | - |
| Zhang et al. [41] | Two methods based on spilt-sample classification | Built by the authors | - |
| Vijayanand et al. [42] | GA-based FS | ADFA-LD CIC-IDS2017 | 99.95 99.39 |
| Alfoudi et al. [43] | 2D-DWTPP | Built by the authors | 98.65 |

## CONCLUSION:

In conclusion, this article reviewed 30 effective techniques leveraging various machine learning and optimization processes in intrusion detection systems. All mechanisms are classified into eight categories according to the primary offered solutions. The suggested analysis in this survey focused on accuracy as the primary criteria. We also checked for processing time and were disappointed to discover that they lacked any system performance data, including processing time. By and large, all groups demonstrated superior skills in terms of the accuracy measure. Additionally, we examined how machine learning methods may be used to cybersecurity and other security-related challenges. In terms of present research, conventional security solutions have garnered considerable attention, whereas security systems based on machine learning techniques have received less attention. We've reviewed pertinent security research for each widely used technique. This article will offer an overview of the conceptualization, understanding, modeling, and reasoning processes involved in cybersecurity data science.